\documentclass[11.pt]{article}

\usepackage{latexsym}
\usepackage{epsfig}
\usepackage{color}
\usepackage{caption2}
\def\R{ {\rm R \kern -.31cm I \kern .15cm}}
\def\C{ {\rm C \kern -.15cm \vrule width.5pt \kern .12cm}}
\def\Z{ {\rm Z \kern -.27cm \angle \kern .02cm}}
\def\N{ {\rm N \kern -.26cm \vrule width.4pt \kern .10cm}}
\def\1{{\rm 1\mskip-4.5mu l} }
\def\lsim{\raise0.3ex\hbox{$<$\kern-0.75em\raise-1.1ex\hbox{$\sim$}}}
\def\gsim{\raise0.3ex\hbox{$>$\kern-0.75em\raise-1.1ex\hbox{$\sim$}}}
\def\noi{\noindent}

\def\beq{\begin{equation}}   \def\eeq{\end{equation}}
\def\bea{\begin{eqnarray}}  \def\eea{\end{eqnarray}}

\def\noi{\noindent}

\DeclareGraphicsExtensions{.eps}

\def\lsim{\raise0.3ex\hbox{$<$\kern-0.75em\raise-1.1ex\hbox{$\sim$}}}
\def\gsim{\raise0.3ex\hbox{$>$\kern-0.75em\raise-1.1ex\hbox{$\sim$}}}

\begin{document}

\title{\bf Proton-proton multiplicity distributions at LHC and the Pomeron intercept}

\vskip 8. truemm
\author{\bf A. Capella$^1$ and E. G. Ferreiro$^2$}
\vskip 5. truemm

\date{}
\maketitle

\begin{center}
\small{
  $^1$ Laboratoire de Physique Th\'eorique\footnote{Unit\'e Mixte de
    Recherche UMR n$^{\circ}$ 8627 - CNRS}, Universit\'e de Paris XI,
  B\^atiment 210, \\
  91405 Orsay Cedex, France
  \par \vskip 3 truemm
   $^2$ Departamento de F{\'\i}sica de Part{\'\i}culas and IGFAE, Universidad de
  Santiago de Compostela, \\
  15782 Santiago de Compostela, Spain}
\end{center}
\vskip 5. truemm

\begin{abstract}
We compute the proton-proton multiplicity distributions at LHC energies in the framework of a multiple scattering model assuming a Poisson distribution for each inelastic collision. Multiple scattering is essential to broaden the multiplicity distribution. We obtain approximate KNO scaling for small pseudo-rapidity intervals ($|\eta | < 0.5$) and sizable KNO scaling violations for larger ones, in agreement with experiment.
\end{abstract}
\vskip 3 truecm

\section{Introduction}
\hspace*{\parindent}
Multiplicity distributions in proton-proton collisions have been measured by the CMS collaboration at $\sqrt{s} = 900$~GeV, 2.36~TeV and $7$~TeV for central pseudo-rapidity intervals $|\eta | < \eta_0$ with $\eta_0 = 0.5$, $1.5$ and $2.4$ \cite{1r}. The measured rapidity distributions are much broader than Poisson distributions and show approximate Koba-Nielsen-Olssen (KNO) scaling \cite{2r} for the smallest pseudo-rapidity interval $\eta_0 = 0.5$ with increasingly larger scaling violations as the length of the interval increases. For $\eta_0 = 2.4$ the scaling violation takes place for $z \equiv n/\left<n\right> \geq 3$. Here $n$ is the event charged multiplicity and $\left<n\right>$ its average value in the considered $\eta$ interval. 

Similar features have been observed at lower energies, between SPS and $p\bar{p}$ collider \cite{3r}. Actually, in this energy range the scaling violations for $\eta_0 = 2.5$ start earlier ($z \geq 2$) and are numerically larger than in the LHC range. These features were well described in the framework of the Dual parton model (DPM) \cite{4r} and Quark gluon string model (QGSM) \cite{5r}. DPM and QGS are multiple-scattering models in which each individual inelastic collision is the superposition of two strings and the weights of the various multiple-scattering contributions are given by a quasi-eikonal model (or perturbative reggeon field theory). One assumes a Poisson distribution for each string for fixed values of the string ends. The broadening of the distribution is due both to the fluctuation in the number of strings and to the fluctuation of the string ends. For energies in the LHC range and pseudo-rapidity intervals of limited length as the ones discussed here, the effect of the fluctuations of the string ends is negligibly small and DPM reduces to an ordinary multiple scattering model with identical multiplicities in each individual scattering \cite{6r}. The broadening of the multiplicity distributions is then entirely due to the fluctuation in the number of inelastic collisions.  

\section{The Model}
\hspace*{\parindent}
Let us consider the charged particle per pseudo-rapidity unit $dN^{pp}/d\eta = (d\sigma^{pp}/d\eta )/\sigma^{pp}_{ND}$ where the numerator is the charged single particle inclusive distribution and $\sigma^{pp}_{ND}$ the non-diffractive $pp$ cross-section. At mid-rapidities and high energies, with identical multiplicities in each individual scattering, we have 
\beq
\label{1e}
{dN^{pp} \over d\eta} = {1 \over \sigma^{pp}_{ND}} \ {d\sigma^{pp} \over d\eta} = {1 \over \sum\limits_{k\geq 1} \sigma_k} \ \sum_{k \geq 1}\  \sigma_k \ k\  {dN_0^{pp} \over d\eta} = \ \left<k\right> {dN_0^{pp} \over d\eta} \ .
\eeq

\noi Here $\left<k\right>$ is the average number of inelastic collisions and $dN_0^{pp} /d\eta$ the charged multiplicity (per pseudo-rapidity unit) in an individual collision. Note that the general formula \cite{4r} for $dN^{pp}/d\eta$ in DPM reduces to the simple expression (\ref{1e}) when all string contributions (and therefore all individual scatterings) are identical. 
This turns out to be the case at LHC energies for the small central rapidity intervals under
consideration.
$dN_0^{pp}/d\eta$ is then equal to twice the string multiplicity per pseudo-rapidity unit. This quantity is independent of $s$ at mid-rapidities and high energies.

Let us turn now to the weights $\sigma_k$ for the occurrence of $k$ inelastic collisions and their energy dependence. Following \cite{6r,7r} we use a quasi-eikonal model with exponential residues in $t$ in which diffractive contributions are included as intermediate states in the eikonal model . We have \cite{6r,7r} 
\beq
\label{2e}
\sigma_k (\xi ) = {\sigma_P \over kZ} \left [ 1 - \exp (-Z) \ \sum_{i=0}^{k-1}\  {Z^i \over i !} \right ] \quad (k \geq 1) \ .
\eeq

\noi Here $\xi = \ell n (s/s_0)$ with $s_0 = 1$~GeV$^2$, $\sigma_P = 8 \pi \gamma_P \exp (\Delta \xi )$ and $Z = 2C_E \gamma_P \exp (\Delta \xi )/(R^2 + \alpha '_P \xi )$.

In (\ref{2e}) $\sigma_P$ is the Born term given by  Pomeron exchange with intercept $\alpha_P (0) = 1 + \Delta$. The (non-diffractive) inelastic cross-section in eq. (\ref{1e}) is $\sigma_{ND}^{pp} (\xi ) = \sum\limits_{k \geq 1} \sigma_k (\xi )$. As for the numerator of eq. (\ref{1e}) one obtains using (\ref{2e}) 
\beq
\label{3e}
\sum_{k \geq 1} \ k \ \sigma_k (\xi ) \equiv \sigma_P (\xi ) \ .
\eeq

\noi This is a well known identity known under the name of AGK cancellation \cite{8r}. It implies that all multiple-scattering contributions vanish identically in the single particle inclusive distribution $d\sigma /d\eta$. Only the Born term contribution $\sigma_P$ is left.

The mid-rapidity values of $dN^{pp}/d\eta$ measured by the ALICE \cite{9r} and CMS \cite{10r} collaborations have an $s$-dependence $dN^{pp}/d\eta (\eta^* = 0) \sim s^{0.11}$. With $\sigma^{pp}_{ND}$ behaving approximately as $s^{0.08}$ \cite{11r} we find from eqs.~(\ref{1e}) to (\ref{3e}), $\alpha_P(0) = 1 + \Delta$ with $\Delta \sim 0.19$ \cite{6r}, a value substantially larger than the one usually considered. In the following we take $\Delta = 0.19$. The values of the other parameters in eq.~(\ref{2e}) are \cite{6r}~: $\alpha '_P = 0.25$~GeV$^{-2}$, $R^2 = 3.3$~GeV$^{-2}$, $\gamma_P = 0.85$~GeV$^{-2}$ and $C_E = 1.8$. The parameters $R^2$ and $\alpha '_P$ control the $t$-dependence of the elastic peak and $C_E$ contain the contribution of diffractive intermediate states.
Its value is obtained from the ALICE results for single and double diffractive cross-sections \cite{11r}. The total $pp$ cross-section is given by 
\beq
\label{4e}
\sigma_{tot} (s) = \sigma_P \ f(z/2) \quad , \quad f(z) = \sum_{\ell = 1}^\infty {(-z)^{\ell - 1} \over \ell \ell !}
\eeq

\noi Note that eq.~(\ref{4e}) is obtained neglecting the real part of the scattering amplitude. However, only its imaginary part contributes to the weights $\sigma_k$ ($k \geq 1$) and to $\sigma_{ND}$ --which are needed for the calculation of the multiplicity distribution. \par

The charged multiplicity per unit pseudo-rapidity is given by
\beq
\label{5e}
{dN^{pp} \over d\eta} = \ \left<k\right>\ {dN_0^{pp} \over d\eta} \quad ; \quad \left<k\right> \ = {\sum\limits_{k\geq 1} k \  \sigma_k \over \sum\limits_{k\geq 1} \sigma_k} = {\sigma_P \over \sigma_{ND}^{pp}} \ .
\eeq

\noi As shown in \cite{6r} the model reproduces the energy dependence of $dN^{pp}/d\eta$ as well as its absolute values with $dN_0^{pp}/d\eta (\eta^* = 0) = 1.5$ --consistent with twice the string multiplicity value. \par

The results for $\sigma_{ND}^{pp}$ and $dN^{pp}/d\eta (\eta^* = 0)$ at various energies are given in ref. \cite{6r}. Since they are needed in the calculation of the multiplicity distribution we have listed them, together with $\sigma_{tot}^{pp}$, in Table 1.
\begin{table}[htb!]
\begin{center}\setlength{\arrayrulewidth}{1pt}
\caption{Values of the total and non-diffractive $pp$ cross-sections and the 
charged particle pseudo-rapidity densities in the central rapidity region for the energy range between 200 and 50000 GeV.}
\label{tab1}
\vskip 0.75cm
\begin{tabular}{cccccccc}
\hline\hline
 $\sqrt{s}$ (GeV) & $dN^{pp}/d\eta (y^*=0)$ & $\sigma^{pp}_{ND}$ (mb) & $\sigma^{pp}_{tot}$ (mb) \\
\hline
200 &  2.99 & 31.22 & 41.62 \\
540 & 3.50 & 38.97 & 54.39 \\
900 & 3.82 & 43.33 & 61.85 \\
1800 & 4.34 & 49.64 & 72.93 \\
2760 & 4.71 & 53.77 & 80.27 \\
5500 & 5.42 & 60.78 & 92.92 \\
7000 & 5.70 & 63.33 & 97.57 \\
14000 & 6.61 & 70.99 & 111.57 \\
50000 & 8.80 & 86.19 & 139.67 \\
\hline\hline
\end{tabular}
\end{center}
\end{table}

\section{Multiplicity distributions}
\hspace*{\parindent}
We turn next to the charged particle multiplicity distributions. The particle production in one-string process is expected to correspond to independent emission of clusters, described by a Poisson distribution. Since the convolution of Poisson distributions is also Poissonian, the multiplicity distribution for a process with $k$ inelastic collisions will be given by a Poisson distribution --with average multiplicity equal to $k$ times the one for a single collision, (or $2k$ times the one of an individual string) 
\beq
\label{6e}
P_{n_c}^{(k)} = e^{-k\left<n_c\right>_0} {(k \left<n_c\right>_0 )^{n_c} \over n_c !} \ .
\eeq

\noi Here $n_c$ is the number of emitted clusters and $\left<n_c\right>_0 = \left<n\right>_0/k$ its average multiplicity in a single inelastic collision. The latter is equal to the average multiplicity of charged particles $\left<n\right>_0$ divided by the average cluster multiplicity $K$, with
\beq
\label{7e}
\left<n\right>_0 \ = \int_{\eta - \eta_0}^{\eta + \eta_0} d\eta\  {d N_0^{pp} \over d\eta} \sim 2\eta_0 \ {dN_0^{pp} \over d \eta} (\eta^* = 0) = 3 \eta_0
\eeq

\noi valid at high energy for the relatively small values of $\eta_0$ under consideration.

The cluster multiplicity distribution is then given by
\beq
\label{8e}
P_{n_c} = {1 \over \sigma_{ND}^{pp}} \ \sum_{k\geq 1} \ \sigma_k \ P_{n_c}^{(k)} \ .
\eeq

\noi Eq.~(\ref{6e}) is strictly valid only for infinitely narrow clusters. However it can be shown to be also true to a high degree of accuracy for clusters decaying according to a Poisson law \cite{12r}. A value of the average cluster multiplicity $K \sim 1.4$ has been obtained in a Monte-Carlo simulation where the clusters are identified with a realistic mixture of directly produced particles and known resonances \cite{13r}. The same value $K = 1.4$ allows to describe the charged multiplicity distributions in $e^+e^-$ and $\ell p$ scattering \cite{12r}. The same value of $K$ is also needed in DPM to reproduce multiplicity distributions, long-range and short-range correlations in $\bar{p}p$ collisions up to $\sqrt{s} = 540$~GeV \cite{4r}. As discussed in \cite{14r}, multiple scattering is responsible for long-range rapidity correlation which increase with $\sqrt{s}$. 

\section{Numerical results}
\hspace*{\parindent}
In the approach described above the $pp$ multiplicity distributions can be computed with no extra parameters. It is convenient to plot them in the KNO form (i.e. $\left<n\right>\ P_n$ versus $z = n/\left<n\right>$). In this case the multiplicity distribution of final particles is identical to the one of clusters, i.e.
\beq
\label{9e}
\psi (z) = \ \left<n\right>\ P_n = \ \left<n_c\right> \ P_{n_c}
\eeq

\noi where $z = n/\left<n\right> = n_c/\left<n_c\right>$. Here $n$ is the number of emitted charged particles and $\left<n\right>$ its average value $\left<n\right> = \left<k\right> \left<n\right>_0 = 3\eta_0 \left<k\right>$. Also $n_c = n/K$ and $\left<n_c\right> = \left<n\right>/K$.

The results for $\psi (z)$ at two energies ($\sqrt{s} = 900$~GeV and 7 TeV) and two values of $\eta_0$ ($\eta_0 = 0.5$ and 2.4) 
are given in Fig.~1 and compared with CMS data \cite{1r}. The corresponding predictions at 14 TeV are also shown.
We see that, in agreement with CMS data, the model has approximate KNO scaling
between 900 GeV and 7 TeV for the small pseudo-rapidity interval $\eta_0=0.5$ 
--up to a large value of $z$, $(z \sim 6)$ where $\psi(z)$ has decreased by three orders of magnitude.
For $\eta_0=2.4$, the KNO scaling violation starts much earlier $(z \sim 2.5)$ and the multiplicity distributions in
KNO form get broader with increasing energy.\par
\begin{figure*}[htb!]
\begin{center}
\begin{flushleft}
\includegraphics[width=0.5\textwidth]{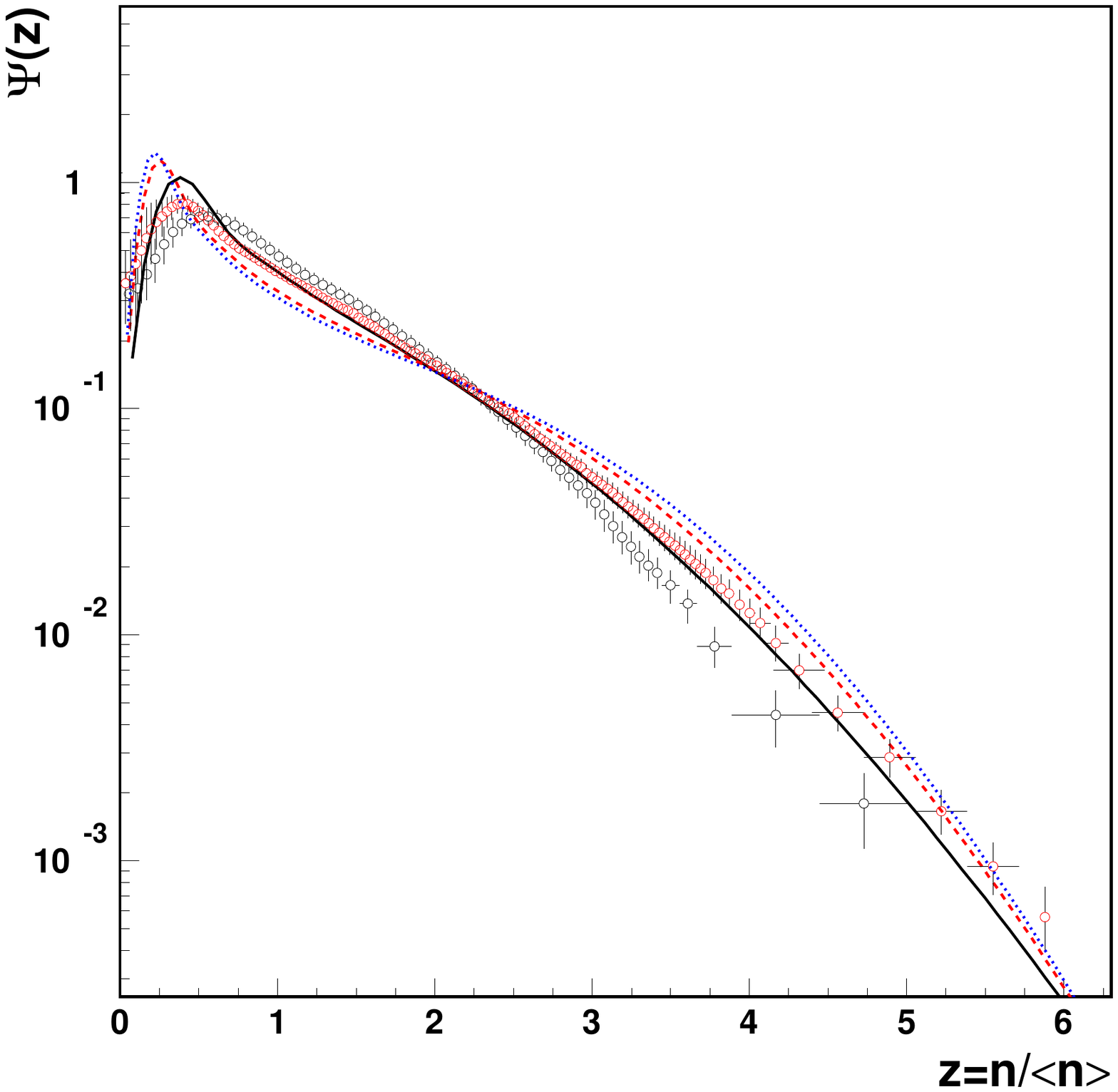}
\end{flushleft}
\begin{flushright}
\vskip -12.42 cm
\includegraphics[width=0.5\textwidth]{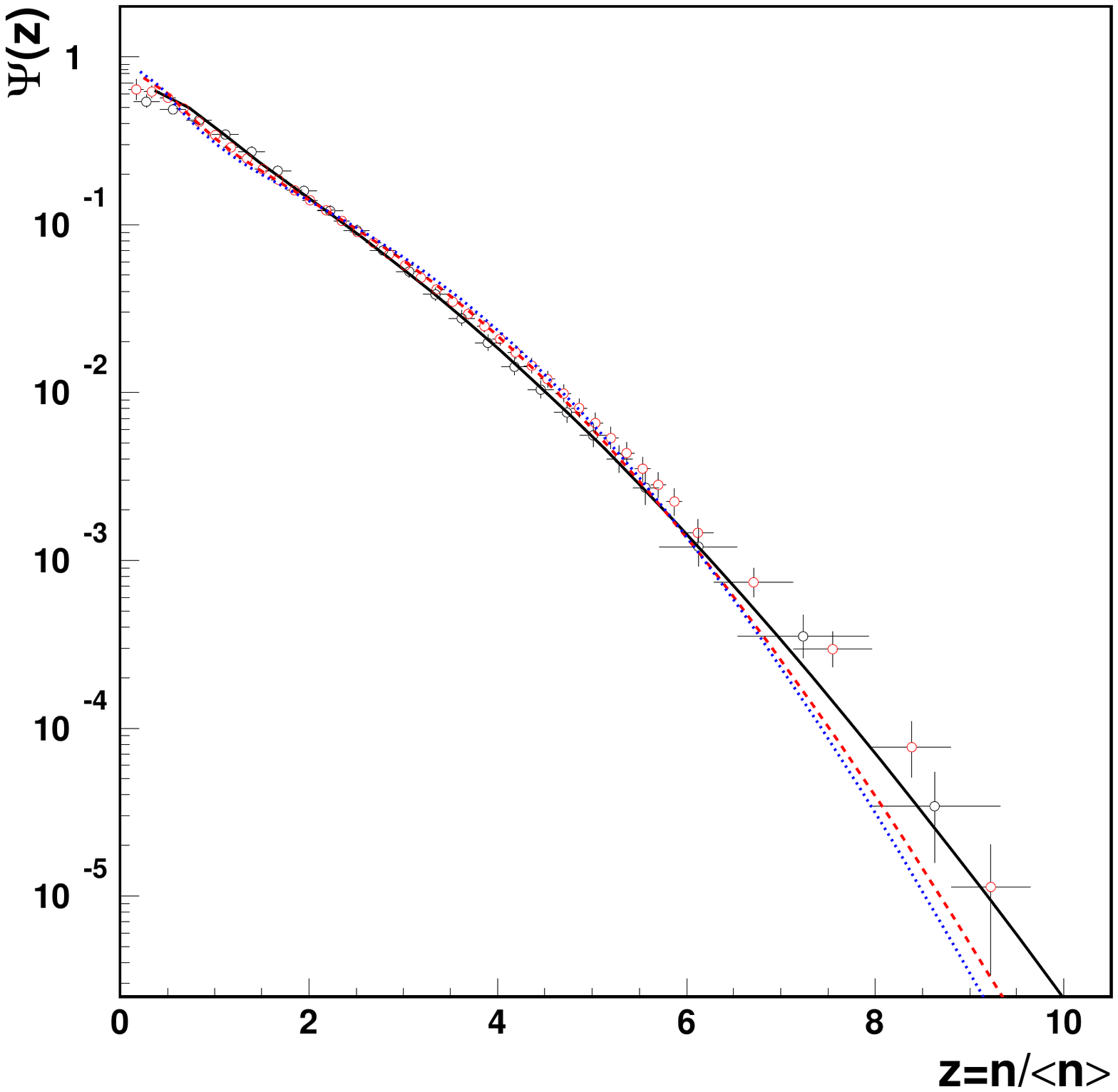}
\end{flushright}
\end{center}
\vskip -2cm
\caption{The charged hadron multiplicity distributions in KNO form at $\sqrt{s}=$ 0.9 TeV (black solid line) and 7 TeV
(red dashed line) TeV in two pseudorapidity intervals, $|\eta| < 2.4$ (left) and $|\eta| < 0.5$ (right) compared to experimental data \cite{1r}. Predictions at 14 TeV (blue pointed line) are also plotted.} 
\label{fig1}
\end{figure*}

%

\section{Conclusions}
\hspace*{\parindent}
We have computed $pp$ multiplicity distributions at LHC in the framework of a
multiple-scattering model (DPM). Multiple-scattering models do not obey KNO
scaling. Indeed, the multiple scattering contributions, which give rise to
long-range rapidity correlations, become increasingly important when $s$
increases and since they contribute mostly to high multiplicities they lead
to KNO multiplicity distributions that get broader as $s$ increases. On the
other hand, the Poisson distributions in the individual scatterings lead to
short-range rapidity correlations and give rise to KNO multiplicity
distributions that get narrower with incresing $s$. Due to the interplay of
these two components the energy dependence of the KNO multiplicity
distributions (or of its normalized moments) depends crucially on the size
of the rapidity interval \cite{16r}. For large rapidity intervals the
multiple-scattering effect dominates and KNO multiplicity distributions get
broader with increasing $s$. For small intervals the effect of the short-range
component increases leading to approximate KNO scaling, up to $z \sim 6$.
We have shown that the above features are maintained  up to the highest LHC
energy and that for a given pseudo-rapidity interval ($\eta_0=2.4$) the rise of
the KNO tail starts at a value of $z$ that increases with energy.

\section*{Acknowledgments}
\hspace*{\parindent}
It is a pleasure to thank Andr\'e Krzywicki for interesting discussions and comments on KNO scaling.
This work is partially supported by MINECO/IN2P3
(AIC-D-2011-0740).

\end{document}